\documentclass[aps,prl,twocolumn,showpacs]{revtex4-1}
\usepackage{amsmath}
\usepackage{amssymb}
\usepackage{graphicx}
\usepackage{epstopdf}

\begin{document}
\title{A Realization of the Haldane-Kane-Mele Model in a System of Localized Spins}

\author{Se Kwon Kim}
\author{H\'{e}ctor Ochoa}
\author{Ricardo Zarzuela}
\author{Yaroslav Tserkovnyak}
\affiliation{Department of Physics and Astronomy, University of California, Los Angeles, California 90095, USA}

\date{\today}

\begin{abstract}
We study a spin Hamiltonian for spin-orbit-coupled ferromagnets on the honeycomb lattice. At sufficiently low temperatures supporting the ordered phase, the effective Hamiltonian for magnons, the quanta of spin-wave excitations, is shown to be equivalent to the Haldane model for electrons, which indicates the nontrivial topology of the band and the existence of the associated edge state. At high temperatures comparable to the ferromagnetic-exchange strength, we take the Schwinger-boson representation of spins, in which the mean-field spinon band forms a bosonic counterpart of the Kane-Mele model. The nontrivial geometry of the spinon band can be inferred by detecting the spin Nernst effect. A feasible experimental realization of the spin Hamiltonian is proposed.
\end{abstract}


\maketitle
\emph{Introduction.}| Electronic systems with spin-orbit coupling (SOC) can exhibit spin Hall effects, in which a longitudinal electric field generates a transverse spin current and vice versa \cite{*[][{, and references therein.}] SinovaRMP2015}. In particular, \textcite{KanePRL2005} showed that a single layer of graphene has a topologically nontrivial band structure with an SOC-induced energy gap, which gives rise to a quantum spin Hall effect characterized by helical edge states. This identification of graphene as a quantum spin Hall insulator has served as a starting point for the search for other topological insulators \cite{*[][{, and references therein.}] HasanRMP2010, *[][{, and references therein.}] QiRMP2011}. 

SOC magnets with no charge degrees of freedom can also exhibit various Hall effects \cite{HaldanePRB1995, *FujimotoPRL2009, *OnoseScience2010, *ZhangPRB2013, *ShindouPRB2013, *ShindouPRB2013-2, *MatsumotoPRB2014, *MookPRB2014, *HirschbergerPRL2015, KatsuraPRL2010, LeePRB2015, MatsumotoPRL2011, *MatsumotoPRB2012}. By exploiting the ubiquitous spin-heat interactions \cite{*[][{, and references therein}] BauerNM2012}, thermal Hall effect, in which a longitudinal temperature gradient induces a transverse heat current, has been used to probe SOC in such insulating magnets. For ordered mangets, thermal Hall effects are often accounted for by geometrically nontrivial band structures of magnons, quanta of spin-wave excitations. For example, \textcite{MatsumotoPRL2011} showed that certain engineered thin-film ferromagnets can have chiral edge states of magnetostatic spin waves (that are associcated with topologically nontrivial bulk band structures) and thus exhibit a thermal Hall effect. For magnets that are disordered due to either thermal or quantum fluctuations, thermal Hall effects have been predicted by using the Schwinger boson or fermion representation of spins \cite{ArovosPRB1988, *AuerbachPRL1988} that do not need a putative symmetry-breaking underlying state \cite{KatsuraPRL2010, LeePRB2015, OkamotoPRB2016}.

In this Letter, we propose a simple spin Hamiltonian for SOC ferromagnets on the honeycomb lattice, in which we can find the bosonic counterparts of both the Haldane \cite{HaldanePRL1988} and the Kane-Mele model \cite{KanePRL2005}. In the ordered phase supported at sufficiently low temperatures, we show that the effective Hamiltonian for magnons is equivalent to the Haldane model \cite{HaldanePRL1988}. For elevated temperatures, where the system is disordered, we take an alternative Schwinger-boson (or bosonic-spinon) representation of spins, in which the mean-field spinon band is identified as a bosonic counterpart of the Kane-Mele model \cite{KanePRL2005}. The nontrivial geometry of the spinon band gives rise to the spin Nernst effect, in which a longitudinal temperature gradient generates a transverse spin current \cite{BauerNM2012}. Lastly we propose a feasible way to realize the underlying Hamiltonian.

\begin{figure}
\includegraphics[width=0.95 \columnwidth]{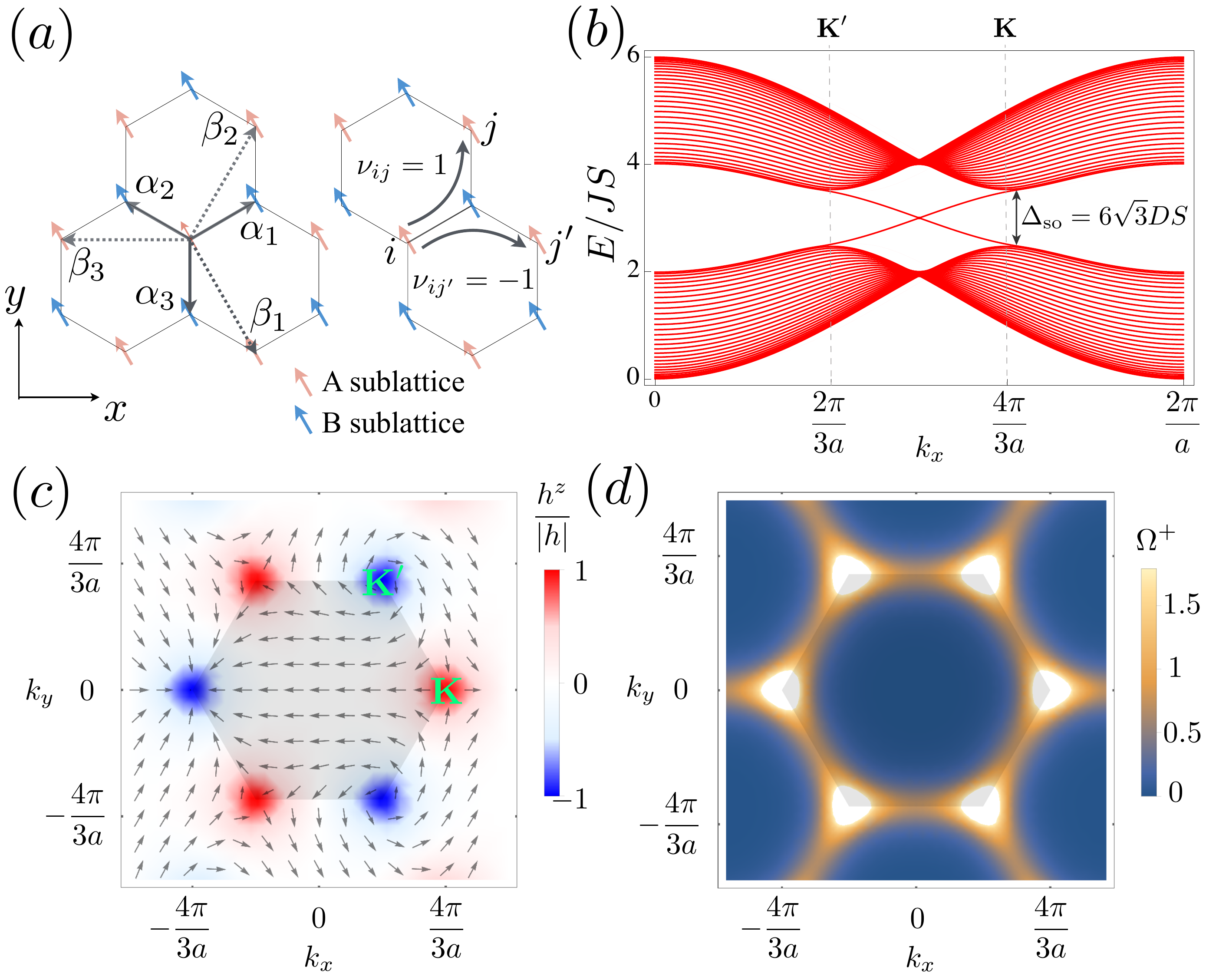}
\caption{(color online) (a) The honeycomb lattice structure and the relative sign $\nu_{ij}$ of the DM interaction. (b) One-dimensional projection of the magnon bands [Eq.~(\ref{eq:E})], which are calculated with a ribbon geometry with zig-zag terminations and 30 unit cells width. (c) The direction of the vector field $\mathbf{h}(\mathbf{k})$ [Eq.~(\ref{eq:h})]. (d) The Berry curvature of the upper band, $\Omega^+$. For (b)-(d), the parameters $D = 0.1 J$ and $B = 0$ are used. The shaded honeycomb in (c) is the first Brillouin zone. Two Dirac points, $\mathbf{K}$ and $\mathbf{K}'$ are denoted in (b) and (c). See the main text for detailed discussions.}
\label{fig:fig1}
\end{figure}

\emph{Model.}|We consider a ferromagnetic material with SOC whose localized spins are arranged on a honeycomb lattice. The corresponding model Hamiltonian reads
\begin{equation}
H = - J\sum_{\langle i, j \rangle} \mathbf{S}_i \cdot \mathbf{S}_j -K\sum_{\langle i, j \rangle} S_{i}^{z}S_{j}^{z}+ D\hspace{-0.1cm}\sum_{\langle \langle i, j \rangle \rangle}\hspace{-0.1cm} \nu_{ij} \hat{\mathbf{z}}\cdot (\mathbf{S}_i\hspace{-0.01cm}\times\hspace{-0.01cm} \mathbf{S}_j),
\label{eq:H}
\end{equation}
where the first and second terms represent the isotropic Heisenberg interaction ($J > 0$) and the Ising interaction between nearest neighbors, respectively \footnote{Each nearest-neighbor pair $\protect\langle i, j \protect\rangle$ contributes only once to the summation in the first and the second term; so does each next-nearest-neighbor pair $\protect\langle \protect\langle i, j \protect\rangle \protect\rangle$ to the third term.}. The third term is the Dzyaloshinskii-Moriya (DM) interaction \cite{DzyaloshinskiiJPCS1958, *MoriyaPR1960} between next-nearest neighbors, where the constants $\nu_{ij} = - \nu_{ji} = \pm 1$ characterize the dependence of the interaction on the relative position of two next-nearest spins [Fig.~\ref{fig:fig1}(a)]. Notice that Eq.~(\ref{eq:H}) represents the minimal Hamiltonian describing the above interactions that are invariant under $D_{6h}$ point-group symmetry of the lattice with concurrent spatial and spin rotations. For the particular experimental realization that we propose in this Letter (see below), the Ising contribution $\propto K$ in Eq.~(\ref{eq:H}) can be safely neglected compared to the other terms. As we are interested in the topological properties of the model for both ordered and disordered phases, we introduce an external magnetic field applied along the $z$ direction [and therefore a Zeeman coupling term $- B \sum_i S_i^z$ in Eq.~(\ref{eq:H})] to stabilize the ferromagnetic ground state. We shall denote by $d$ and by $a = \sqrt{3} d$ the distances between nearest neighbors and next-nearest neighbors, respectively. 

%The Hamiltonian $H$ respects the following symmetries similar to the Kane and Mele's tight-binding Hamiltonian for electrons in graphene, specifically Eq.~(6) in Ref.~\cite{KanePRL2005}: 1) the translational and point-group symmetry of the honeycomb lattice, 2) the global spin-rotational symmetry about the $z$ axis, and 3) the time reversal symmetry in the absence of an external field. Note also, in both Hamiltonians, SOC breaks the otherwise-respected sublattice symmetry, $\mathcal{A} \leftrightarrow \mathcal{B}$ in a spin-dependent way.

\emph{Magnon picture.}|The uniform state $\mathbf{S}_i \equiv S \hat{\mathbf{z}}$ represents the classical ground state of the model Hamiltonian \eqref{eq:H} for $B \ge 0$ and $D < J / \sqrt{3}$. Application of the Holstein-Primakoff transformation $S_i^+ = S_i^x + i S_i^y = (2 S - n_i)^{1/2} d_i$, $S_i^- = (S_i^+)^\dagger$, and
$S_i^z = S - n_i$ with $n_i = d_i^\dagger d_i$ yields the following effective magnon Hamiltonian
\begin{align}
\label{eq:Hm}
H_\text{m} = 	& (3 J S + B) \sum_i d_i^\dagger d_i - J S \sum_{\langle i, j \rangle} \left[ d_i^\dagger d_j + \text{h.c.} \right] \\
			& \hspace{0.5cm}- D S \sum_{\langle \langle i, j \rangle \rangle} \left[ i \nu_{ij} d_i^\dagger d_j + \text{h.c.} \right],\nonumber
\end{align}
up to second order in the magnon operators $d_{i}$ and $d_{i}^{\dagger}$. In this approximation, the Hamiltonian reduces to the Haldane model \cite{HaldanePRL1988}.

The topological features of the magnon bands can be readily captured in the momentum representation. Let $\Psi_\mathbf{k} = (a_\mathbf{k}, b_\mathbf{k})$ be the spinor operators in the Fourier space, where $a$ and $b$ represent magnon annihilation operators on the sublattices $\mathcal{A}$ and $\mathcal{B}$, respectively. Fourier transform of the Hamiltonian \eqref{eq:Hm} then reads
\begin{equation}
H_\text{m} = \sum_{\mathbf{k} \in \text{B.Z.}} 
\Psi_\mathbf{k}^\dagger
\left[ (3 J S + B) I + \mathbf{h} (\mathbf{k}) \cdot \boldsymbol{\tau} \right]
\Psi_\mathbf{k},
\end{equation}
where $\boldsymbol{\tau}$ is a pseudovector of the Pauli matrices and
\begin{equation}
\mathbf{h}(\mathbf{k}) = 
\sum_{j}
\begin{pmatrix}
	- J S \cos[\mathbf{k} \cdot \boldsymbol{\alpha}_j] \\
	J S \sin[\mathbf{k} \cdot \boldsymbol{\alpha}_j] \\
	2 D S \sin[\mathbf{k} \cdot \boldsymbol{\beta}_j]
\end{pmatrix}
=
    \begin{cases}
      3 \sqrt{3} D S \hat{\mathbf{z}} & \mathbf{k}=\mathbf{K}\\
     -3 \sqrt{3} D S \hat{\mathbf{z}} & \mathbf{k}=\mathbf{K}'
    \end{cases},
\label{eq:h}
\end{equation}
where $\boldsymbol{\alpha}_{i}$ and $\boldsymbol{\beta}_{i}$ are defined in Fig.~\ref{fig:fig1}(a), $\mathbf{K} \equiv (4 \pi / 3 a, 0)$, and $\mathbf{K}' \equiv (2 \pi / 3 a, 2 \pi / \sqrt{3} a)$. The dispersions of the upper and the lower energy band are given by 
\begin{equation}
E_{\textrm{m}}^{\pm} (\mathbf{k}) = 3 J S + B \pm \left| \mathbf{h} (\mathbf{k}) \right| \, .
\label{eq:E}
\end{equation}
In the absence of SOC ($D = 0$), the upper and the lower band meet at two points, $\mathbf{K}$ and $\mathbf{K}'$, forming linearly-dispersed bands \cite{FranssonarXiv2015, OwerrearXiv2016-2}. SOC opens an energy gap $\Delta_\text{so} = 6 \sqrt{3} D S$ at these points, making the band structure topologically nontrivial \cite{KanePRL2005}. Figures~\ref{fig:fig1}(b) and (c) show the one-dimensional projection of the magnon bands $E^\pm (\mathbf{k})$ and the direction of $\mathbf{h} (\mathbf{k})$, respectively, for the values $D = 0.1 J$ and $B = 0$ \footnote{To be precise, Fig.~\ref{fig:fig1}(b) is obtained by combining the plots of $E^\pm(k_x, k_y)$ as a function of $k_x$ for all $k_y$.}. 

The Berry curvatures of the upper and the lower magnon bands can be calculated according to the formula $\Omega_{\textrm{m}}^{\pm} = \mp \hat{\mathbf{n}} \cdot (\partial_{k_x} \hat{\mathbf{n}} \times \partial_{k_y} \hat{\mathbf{n}}) / 2$, where $\hat{\mathbf{n}}$ is the unit vector along $\mathbf{h}$ \cite{XiaoRMP2010}. Figure~\ref{fig:fig1}(d) plots the Berry curvature of the upper band. Notice that the Berry curvature is large around the corners of the Brillouin zone, $\mathbf{K}$ and $\mathbf{K}'$, where the vector $\mathbf{n}$ exhibits nontrivial topological textures that wrap a half of the unit sphere. The Chern numbers \cite{ShindouPRB2013-2} of the bands are evaluated as $C^\pm = (1 / 2 \pi) \int_{B.Z.} \Omega_{\textrm{m}}^{\pm}\, d^2k = \pm 1$.

\emph{Spinon picture.}|While the magnon picture is valid at sufficiently low temperatures where the system is ordered, it fails when the system is disordered due to thermal fluctuations. For high temperatures comparable to the exchange strength $J$, the Schwinger-boson representation of spins \cite{ArovosPRB1988} provides an alternative approach to study the topological features of the spin system. The corresponding transformation reads $S^+_i = c_{i, \uparrow}^\dagger c_{i, \downarrow}$, $S_i^- = c_{i, \downarrow}^\dagger c_{i, \uparrow}$ and $S_i^z = (c_{i, \uparrow}^\dagger c_{i, \uparrow} - c_{i, \downarrow}^\dagger c_{i, \downarrow}) / 2$. Here $c_{i, s}$ ($c_{i, s}^\dagger$) represents the annihilation (creation) operator of spin-$1/2$ up- or down bosons at the site $i$, which are referred to as Schwinger bosons or bosonic spinons. The local number constraint, $\sum_{s} c_{i, s}^\dagger c_{i, s} = 2 S$, needs to be imposed to fulfill the spin-$S$ algebra. The Hamiltonian \eqref{eq:H} in the spinon picture reads

%After the local constraint is imposed by using the Lagrangian multiplies $\lambda_i$, the Hamiltonian reads
\begin{align}
H_{\textrm{s}} = 	& - 2 J \sum_{\langle i, j \rangle} \chi_{ij}^\dagger \chi_{ij} - \frac{B}{2} \sum_i \left( c_{i, \uparrow}^\dagger c_{i, \uparrow} - c_{i, \downarrow}^\dagger c_{i, \downarrow} \right)\\
	& - \frac{D}{2} \sum_{\langle \langle i, j \rangle \rangle} i\nu_{ij} \left( \chi_{ij, \uparrow}^\dagger \chi_{ij, \downarrow} - \chi_{ij, \downarrow}^\dagger \chi_{ij, \uparrow} \right) \nonumber\\
	&+ \sum_i \lambda_i \left( c_{i, \uparrow}^\dagger c_{i, \uparrow} + c_{i, \downarrow}^\dagger c_{i, \downarrow} - 2 S \right) \, , \nonumber
\end{align}
up to a constant, where $\chi_{ij, s} = c_{i, s}^\dagger c_{j, s}$ are operators defined for pairs of sites for each spin $s$, $\chi_{ij} = \left( \chi_{ij, \uparrow} + \chi_{ij, \downarrow} \right) / 2$, and $\lambda_i$ is the Lagrange multiplier related to the above holonomic constraint. 

\begin{figure}
\includegraphics[width= \columnwidth]{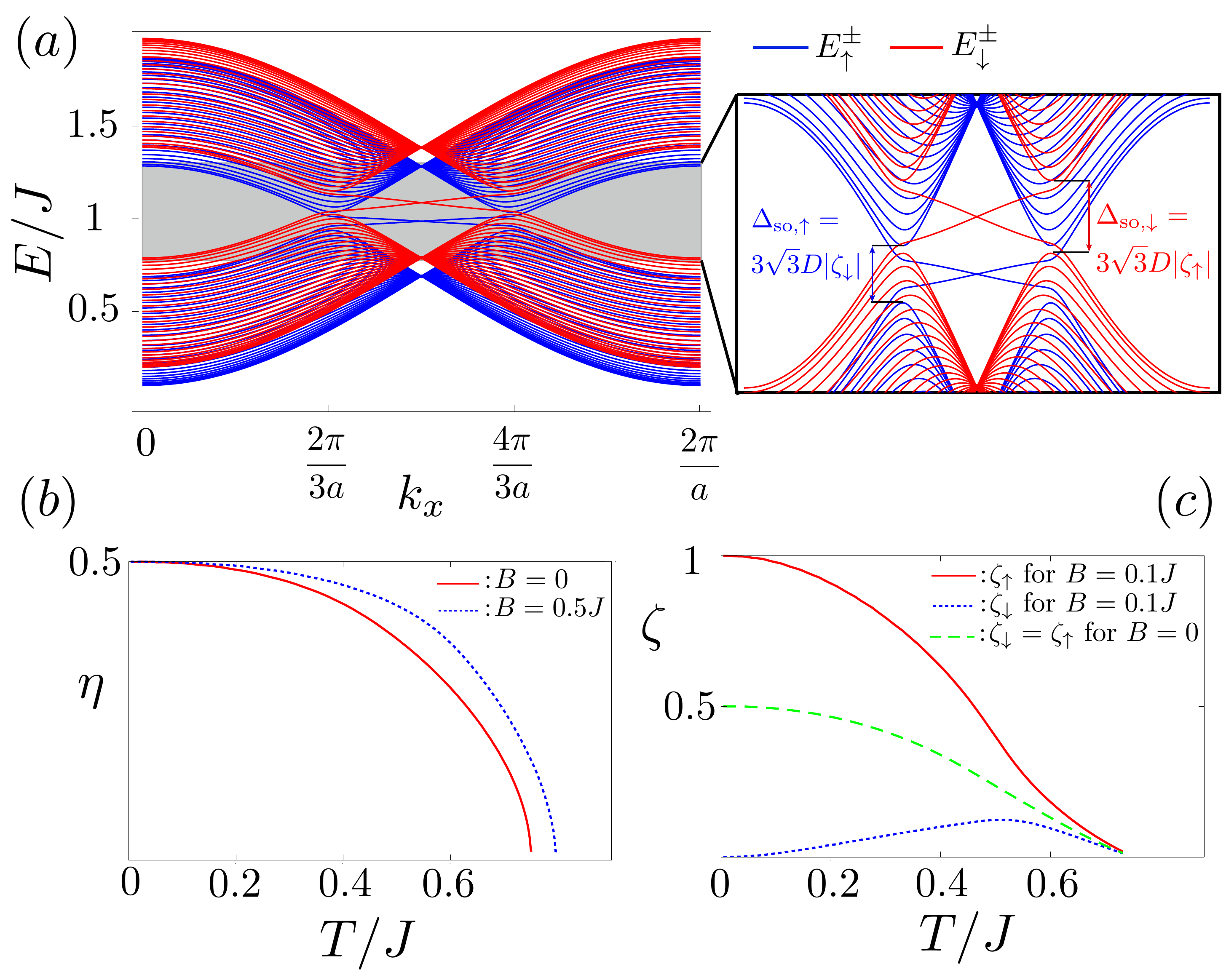}
\caption{(color online) (a) One-dimensional projection of the spinon bands [Eq.~(\ref{eq:Es})] at the temperature $T = 0.6J$ for the parameters $S = 1/2, D = 0.1 J, B = 0.1J$. A zoom of the middle sector is shown in the right. The band structure calculation corresponds to a ribbon geometry with zig-zag terminations and 30 unit cells. (b), (c) The dependence of the mean fields $\eta, \zeta_\uparrow$, and $\zeta_\downarrow$ on the temperature for the parameters $S = 1/2$ and $D = 0.1 J$. See the main text for detailed discussions.}
\label{fig:fig2}
\end{figure}

As the first and third terms are quartic in the spinon operators, we take the mean-field approach \footnote{The mean-field configuration corresponds to the saddle-point solution of the action in the path-integral formulation \cite{ArovosPRB1988}.}. We choose the Hartree-Fock decoupling \cite{Bruus} that retains the symmetries of the original Hamiltonian by conserving the total number of spinons and the $z$-component of the total spin by following Ref.~\cite{LeePRB2015}. First, we use the mean field $\eta = \langle \chi_{ij} \rangle = \sum_s \langle c_{i, s}^\dagger c_{j, s} \rangle$ for nearest neighbors $i \in \mathcal{A}$ and $j \in \mathcal{B}$. While $\eta$ is a complex number generally, we can assume it real by absorbing its factor into the operator $c_{j, s}$. We then make the substitution, $\chi_{ij}^\dagger \chi_{ij} \mapsto \eta \chi_{ij} + \eta \chi_{ij}^\dagger - \eta^2$ in the Hamiltonian. Secondly, for next-nearest neighbors, we define two mean fields for $\chi_{ij, s}$: $\zeta_s = \langle \chi_{ij, s} + \chi_{ji, s} \rangle / 2$ for the symmetric part (with respect to $i \leftrightarrow j$) and $\xi_s = \nu_{ij} \langle \chi_{ij, s} - \chi_{ji, s} \rangle / 2 i$ for the antisymmetric part. The two mean fields are real owing to $\chi_{ji, s} = \chi_{ij, s}^\dagger$. We then perform the necessary substitution. Thirdly, we replace the local Lagrange multipliers $\lambda_i$ by the global one $\lambda$. The resultant mean-field Hamiltonian is given by
\begin{align}
\label{eq:Hs}
H_{\textrm{s}} = & - \eta J \sum_{\langle i, j \rangle, s} \left[ c_{i, s}^\dagger c_{j, s} + \text{h.c.} \right] \\
&+ \frac{D}{2} \sum_{\langle \langle i, j \rangle \rangle, s} \left[i\nu_{ij} s \zeta_{-s}\, c_{i, s}^\dagger c_{j, s} + \text{h.c.} \right] \nonumber\\
&+ \frac{D}{2} \sum_{\langle \langle i, j \rangle \rangle, s} \left[ s \xi_{-s}\, c_{i, s}^\dagger c_{j, s} + \text{h.c.} \right]\nonumber\\
&+ \sum_{i,s}\left(\lambda - s\frac{B}{2}\right) c_{i, s}^\dagger c_{i, s} \, .\nonumber
\end{align}
The mean fields $\eta$ and $\zeta_s$ represent short-ranged spin correlations \cite{SarkerPRB1989}. The first two terms in the above Hamiltonian then correspond to the Kane-Mele model \cite{KanePRL2005}, from which we can infer the nontrivial topology of the spinon-band structure and the existence of edge states for both spin-up and spin-down spinons. As we shall discuss below, the third term of Eq.~(\ref{eq:Hs}) does not affect the topological features of the spinon bands. 

The spinon Hamiltonian in the momentum representation reads 
\begin{equation}
H_{\textrm{s}} = \sum_{\mathbf{k} \in \text{B.Z.},\, s} \Psi_{\mathbf{k}, s}^\dagger \left[ g_s (\mathbf{k}) I + \mathbf{h}_s (\mathbf{k}) \cdot \boldsymbol{\tau} \right] \Psi_{\mathbf{k}, s} \, , 
\end{equation}
where $\Psi_{\mathbf{k}, s} = (a_{\mathbf{k}, s}, b_{\mathbf{k}, s})$ is the spinor of annihilation operators, $g_s (\mathbf{k}) = \lambda - s B / 2 + sD \xi_{-s} \sum_j \cos (\mathbf{k} \cdot \boldsymbol{\beta}_j)$, and
\begin{equation}
\mathbf{h}_s (\mathbf{k}) = 
\sum_{j}
\begin{pmatrix}
	- J \eta \cos[\mathbf{k} \cdot \boldsymbol{\alpha}_j] \\
	J \eta \sin[\mathbf{k} \cdot \boldsymbol{\alpha}_j] \\
	- D s\zeta_{-s} \sin[\mathbf{k} \cdot \boldsymbol{\beta}_j]
\end{pmatrix} \, ,
\label{eq:hs}
\end{equation}
where the vectors $\boldsymbol{\alpha}_i$ and $\boldsymbol{\beta}_j$ are depicted in Fig.~\ref{fig:fig1}(a). The corresponding upper and lower energy bands for each spin $s$ are then given by
\begin{equation}
E^\pm_{s} (\mathbf{k}) = g_s (\mathbf{k}) \pm \left| \mathbf{h}_{s} (\mathbf{k}) \right|. 
\label{eq:Es}
\end{equation}
Notice that the spin-down spinon bands mimic the magnon bands owing to the similarity between the momentum dependence of $\mathbf{h}_s (\mathbf{k})$ [Eq.~(\ref{eq:hs})] and $\mathbf{h} (\mathbf{k})$ [Eq.~(\ref{eq:h})].
 
Self-consistency of the mean-field approach is guaranteed through the equations in momentum space \footnote{Formally, the self-consistency conditions can be obtained by demanding the functional derivative with respect to the mean fields of the free energy $H = H_{\textrm{s}} + 6 J N |\eta|^2 - 6 D N \sum_s s \zeta_s \xi_{-s} - 4 S N \lambda$ to vanish.}:
\begin{align}
\label{eq:scc}
2 S & = \frac{1}{2 N}\sum_{\mathbf{k},\,s}\left[\rho^{-}_{s} (\mathbf{k})+\rho^{+}_{s} (\mathbf{k}) \right] , \\
12 N &=  J\sum_{\mathbf{k},\,s}\frac{\rho^{-}_{s} (\mathbf{k})-\rho^{+}_{s} (\mathbf{k})}{|h_{s}(\mathbf{k})|}\Big|\sum_{i}e^{i\mathbf{k}\cdot\boldsymbol{\alpha_{i}}}\Big|^{2},\nonumber \\
\zeta_{s} &= \frac{1}{6N}\sum_{\mathbf{k}}\left[\rho^{-}_{s} (\mathbf{k})+\rho^{+}_{s} (\mathbf{k}) \right]\Big[\sum_{i}\cos(\mathbf{k}\cdot\boldsymbol{\alpha_{i}})\Big], \nonumber\\
\xi_{s} &= \frac{sD\zeta_{-s}}{6N}\sum_{\mathbf{k}}\frac{\rho^{-}_{s} (\mathbf{k})-\rho^{+}_{s} (\mathbf{k})}{|h_{s}(\mathbf{k})|}\Big|\sum_{i}\sin(\mathbf{k}\cdot\boldsymbol{\alpha_{i}})\Big|^{2},\nonumber
\end{align}
where $\rho^{\tau}_{s} (\mathbf{k}) =\left[ \exp(E^{\tau}_{s} (\mathbf{k}) / T) - 1 \right]^{-1}$ is the Bose-Einstein distribution of spin-$s$ spinons in the $\tau$ band and $N$ is the number of unit cells. Note that the total number of spinons is fixed by the first condition. This enables the Bose condensation of spinons in the limit of zero temperature, which corresponds to magnetic ordering \cite{SarkerPRB1989}. The mean-field spinon bands are obtained by solving self-consistently Eqs.~(\ref{eq:Es}) and (\ref{eq:scc}), which are shown in Fig.~\ref{fig:fig2}(a) at the temperature $T = 0.6 J$ for the values $S = 1/2, D = 0.1 J$, and $B = 0.1 J$. The SOC induces an energy gap $\Delta_{\text{so}, s} = 3 \sqrt{3} D |\zeta_{-s}|$ between the spin-$s$ spinon bands, whose Chern numbers read $C^\pm_{s} = \mp s$. Therefore, the topological nontriviality of the bulk bands for each spin supports the edge states. The thermal dependence of the mean fields $\eta, \zeta_{\uparrow}$ and $\zeta_{\downarrow}$ for the parameters $S = 1/2, D = 0.1J$ is shown in Figs.~\ref{fig:fig2}(b) and (c). The vanishing of these fields at a finite temperature marks a transition between phases with finite (low-$T$) and zero (high-$T$) correlation lengths of spins. Our mean fields represent short-ranged (between 1st and 2nd nearest neighbors) spinon correlation functions, and thus this phase transition is not associated with an onset of long-ranged ordering. The presence of the seriously disordered high-$T$ phase with no correlation between spins is expected to be an artifact of mean-field treatments, i.e., the limit of $N \rightarrow \infty$ in the generalization of the symmetry group from SU(2) to SU(N), as for the case of the pure Heisenberg model ($D = 0$) where it has been shown to disappear after accounting for fluctuations from the mean fields \cite{TchernyshyovNPB2002}.

A connection of the spinon picture to the magnon picture can be established by taking the zero-temperature limit $T \rightarrow 0$, where the spin-down spinon bands are equivalent to the magnon bands and the spin-up spinons form the Bose-Einstein condensation that is the ordered ground state in the magnon picture. To see this, let us apply an external magnetic field $B > 0$, which makes the system completely polarized along the $z$ axis, $\mathbf{S}_i \equiv S \hat{\mathbf{z}}$, as $T \rightarrow 0$. This polarization corresponds in the spinon picture to the Bose-Einstein condensation of spinons into the lowest-energy mode localized at the $\mathbf{k} = 0$ state for the spin-up spinon band \cite{SarkerPRB1989}. The mean fields associated with this polarized state are $\eta = S, \zeta_{\uparrow} = 2 S, \,\zeta_{\downarrow} =\xi_{\uparrow} = \xi_{\downarrow} = 0$ and $\lambda = 3 J S + B / 2$, for which the spin-down spinon bands are equivalent to the magnon bands, $E_{\textrm{m}}^\pm (\mathbf{k}) \equiv E^\pm_{\downarrow} (\mathbf{k})$ \footnote{The agreement between the magnon and the spin-down spinon bands in the zero-temperature limit can be also understood by considering the operator $S_i^-$ in the ground state $\mathbf{S}_i \equiv S \hat{\mathbf{z}}$. Specifically, $S_i^- = d_i^\dagger \sqrt{2S - n_i} \approx \sqrt{2 S} d_i^\dagger$ in the magnon picture, where we used $S \gg n_i$; $S_i^- = c_{i, \downarrow}^\dagger c_{i, \uparrow} \approx \sqrt{2 S} c_{i, \downarrow}^\dagger$ in the spinon picture, where we used $c_{i, \uparrow} \approx \sqrt{2 S}$ in the condensed state \cite{SarkerPRB1989}. In this approximation, the magnon and spin-down spinon operators are equal.}.
.

\begin{figure}
\includegraphics[width=0.95 \columnwidth]{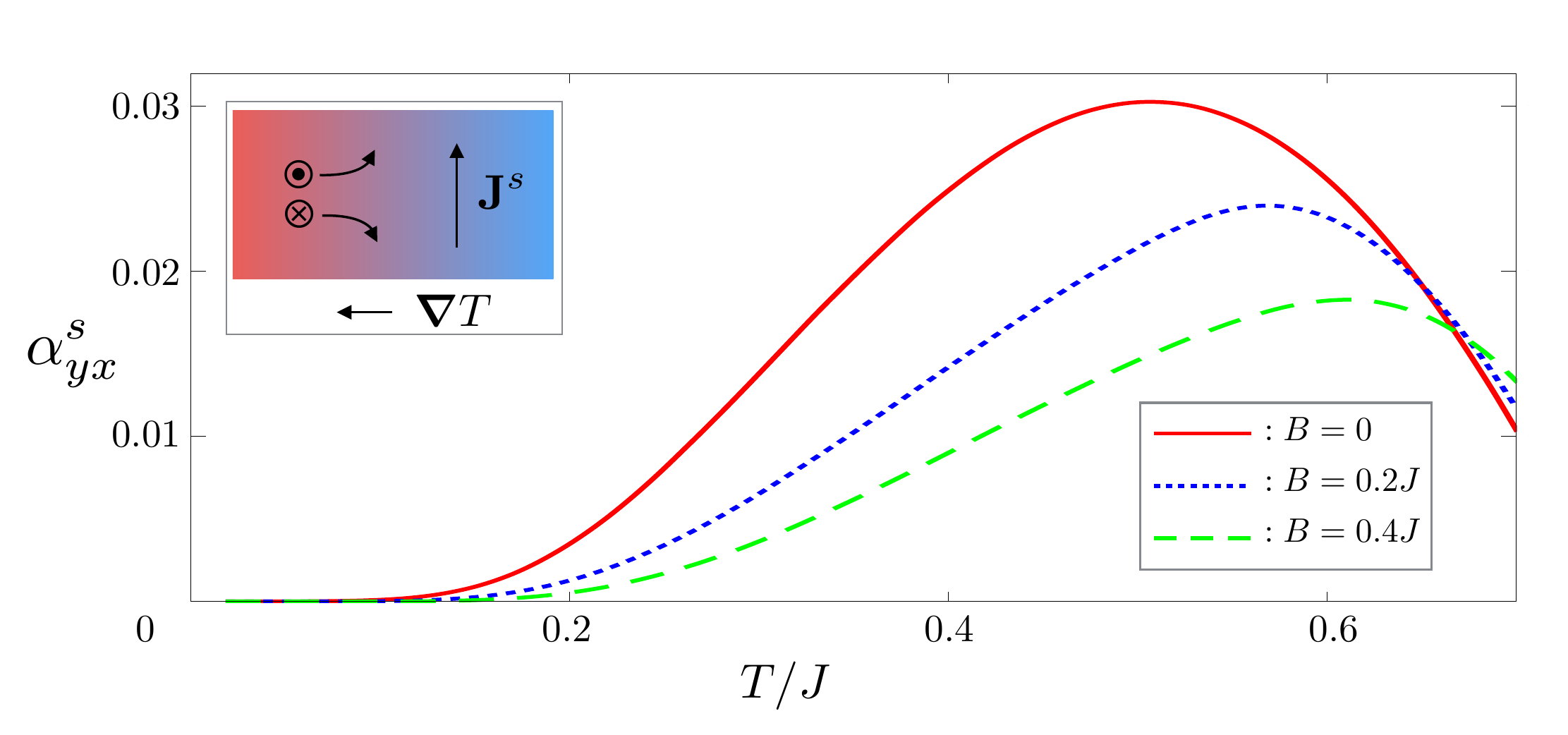}
\caption{(color online) The spin Nernst conductivity $\alpha^s_{yx}$ as a function of temperature. The inset schematically shows a setup of an experiment and motions of spinons therein.}
\label{fig:fig3}
\end{figure}

\emph{Spin Nernst effect.}|Spin-up and spin-down spinons experience opposite Berry curvatures, $\Omega^\pm_\uparrow (\mathbf{k}) \equiv - \Omega^\pm_\downarrow (\mathbf{k})$, in the absence of an external magnetic field. This can induce the spin Nernst effect \cite{BauerNM2012}, in which a transverse spin current is generated by applying a longitudinal temperature gradient, $J^s_y = - \alpha^s_{yx} \partial_x T$. The spinon picture is well suited to compute the spin Nernst conductivity $\alpha^s_{yx}$ due to its applicability over a broad range of temperatures. We use the expression for $\alpha^s_{yx}$ derived in Ref.~\cite{KovalevPRB2016} for the free magnon bands, $\alpha^s_{yx} = - (2 V)^{-1} \sum_{\mathbf{k}, s, \tau} c_1 [ \rho^\tau_s (\mathbf{k}) ] \Omega^\tau_s (\mathbf{k})$, where $V = 3 \sqrt{3} N d^2 / 2$ is the volume of the system and $c_1(x) = (1 + x) \ln (1 + x) - x \ln x$.

Figure~\ref{fig:fig3} shows the thermal dependence of the spin Nernst conductivity $\alpha^s_{yx}$ for the physical parameters $S = 1/2, D = 0.1J$. At zero temperature $\alpha^s_{yx} = 0$ due to the absence of thermal excitations. As the temperature increases, spinons are thermally populated and $\alpha^s_{yx}$ becomes finite. As the temperature approaches the ferromagnetic-exchange strength $J$, the magnitudes of the mean fields start decreasing. The bands thereby flatten more and have smaller Berry curvatures, which in turn results in the suppression of $\alpha^s_{yx}$. Application of a finite magnetic field increases the energies of the spin-down spinons, which in turn decreases the magnitude of the spin Nernst effect. The magnon picture should give similar numerical results for the spin Nernst effect for low temperatures and small magnetic field $T, B \ll J$, owing to the equivalence between the magnon and the spin-down spinon bands in the limit $T \rightarrow 0$ and also the relation between two spinon bands $\Omega^\pm_\uparrow (\mathbf{k}) \equiv - \Omega^\pm_\downarrow (\mathbf{k})$ in the limit $B \rightarrow 0$.

\emph{Discussion}|Although, to the best of our knowledge, the proposed Hamiltonian does not correspond to any existing material, the model may be engineered by depositing magnetic impurities on metals with strong spin-orbit coupling. The minimal Hamiltonian consists of two terms, $\mathcal{H}=\mathcal{H}_{it} +\mathcal{J}\mathbf{S}_i\cdot\mathbf{s}\left(\mathbf{R}_i\right)$. The first term describes the dynamics of surface electrons, whereas the second one describes the coupling between the localized spins $\mathbf{S}_i$ and the spin density of the metal evaluated at the position of the impurities, $\mathbf{R}_i$. Magnetic interactions are mediated by itinerant electrons through the Ruderman-Kittel-Kasuya-Yosida \cite{RudermanPR1954, *KasuyaPTP1956, *YosidaPR1957} interaction. When the system respects both mirror ($z\rightarrow -z$) and C$_{6v}$ point-group symmetries, the effective Hamiltonian reduces to Eq.~\eqref{eq:H}. We provide two exemplary realizations of such system in the Supplementary Material \footnote{See Supplemental Material for details.}. It is worth remarking that the Ising-like coupling $K$ appears as a second order effect in the SOC, which justifies neglecting it over the other first order terms in the SOC.

Breaking the mirror symmetry, e.g., by an external electric field $\propto \hat{\mathbf{z}}$, can generate a DM interaction between the nearest neighbors \cite{FranssonPRL2014},
\begin{equation}
H'=D'\sum_{\left\langle i, j \right\rangle}\left(\hat{\mathbf{z}}\times\boldsymbol{\alpha}_{ij}\right)\cdot\left(\mathbf{S}_i\times\mathbf{S}_j\right) \, .
\end{equation}
The term translates into a Rashba-like hopping term in the spinon mean-field Hamiltonian:
\begin{equation}
H_s' = \frac{D \eta}{4} \sum_{\langle i, j \rangle} \sum_{s,s'} \left[ i \left( \hat{\mathbf{z}}\cdot\boldsymbol{\alpha}_{ij}\times\boldsymbol{\tau} \right)_{s,s'}c_{i,s}^{\dagger}c_{j,s'} + \text{h.c.} \right] \, .
\end{equation}
This term competes with the intrinsic DM interaction in Eq.~(\ref{eq:H}) and can close the topological gaps at the Dirac points if sufficiently strong as in the Kane-Mele model \cite{KanePRL2005}. In the presence of a strong magnetic field $B$, however, the effect of the term is of order $D / B^2 \ll D$ in the perturbative treatment due to the energy separation between the $s=\uparrow$ and $\downarrow$ bands, which allows us to neglect its effect on the gaps $\Delta_{\text{so}, s} \propto D$. %This magnetic-field-induced suppression of the effect of the mirror-symmetry breaking term on the topological gap is not allowed in the original Kane-Mele model which requires the time reversal symmetry.

Another possibility would be using chromium tri-halides like CrBr$_3$, which consist of weakly-coupled ferromagnetic honeycomb layers \cite{*[][{, and references therein.}] DeJonghAP2001}, with the DM interaction induced by the proximity effect with strong spin-orbit coupled materials, e.g., Pt.

%Another possibility would be to consider materials like chromium tri-halides, e.g., CrBr$_3$ or CrCl$_3$, which consist of weakly-coupled ferromagnetic honeycomb layers \cite{*[][{, and references therein.}] DeJonghAP2001}. The interfacial DM interaction can be induced to those materials by the proximity effect with strong spin-orbit coupled materials, e.g., Pt or Ta, which may give rise to the physics described in the Letter.

% In conclusion, we have proposed a simple spin model on the honeycomb lattice, whose properties, including the spin Nernst effect, can be understood in terms of the non-trivial topology of the spectrum of quasiparticle excitations. 
In the spinon picture, we have neglected fluctuations of the Lagrangian multiplier $\lambda_i$ and the bond operators $\chi_{ij, s}$ from their mean-field values, which can be taken into account by, e.g., performing $1/N$ corrections (the mean-field treatment corresponds to generalizing the spin symmetry group from SU(2) to SU(N) and taking $N \rightarrow \infty$ limit) \cite{TrumperPRL1997}. In particular, the phase fluctuations of the bond operators couple to the spinons as the U(1) gauge fields, which has been shown to result in confining the spinons in the ordered phases of some frustrated magnets, e.g., the Heisenberg antiferromagnet on the square lattice \cite{ReadPRL1991, *NgPRB1993}. Investigating effects of the mean-field fluctuations in our spinon picture is a topic for future research.

After the completion of the manuscript, we became aware of recent related works \cite{OwerrearXiv2016-2}, in which the author studied the topological property of the magnon band on the honeycomb lattice and the associated thermal Hall effects. The spinon bands and spin Nernst effects, however, are not discussed in the reference.

\begin{acknowledgments}
We are grateful to Elihu Abrahams, Yong P. Chen, Fenner Harper, and Jing Shi for insightful discussions. This work was supported by the Army Research Office under Contract No. 911NF-14-1-0016, by the U.S. Department of Energy, Office of Basic Energy Sciences under Award No. DE-SC0012190, and by the NSF-funded MRSEC under Grant No. DMR-1420451. RZ thanks Fundaci\'{o}n Ram\'{o}n Areces for the postdoctoral fellowship within the XXVII Convocatoria de Becas para Ampliaci\'{o}n de Estudios en el Extranjero en Ciencias de la Vida y de la Materia.

\end{acknowledgments}

\bibliographystyle{/Users/evol/Dropbox/School/Research/apsrev4-1-nourl}
\bibliography{/Users/evol/Dropbox/School/Research/master}

\end{document}